# Harmony-Search and Otsu based System for Coronavirus Disease (COVID-19) Detection using Lung CT Scan Images

V. Rajinikanth[1], Nilanjan Dey[2], Alex Noel Joseph Raj[3], Aboul Ella Hassanien[4], K.C. Santosh[5], and N. Sri Madhava Raja[1]

[1] Department of Electronics and Instrumentation Engineering, St. Joseph's College of Engineering, Chennai 600119, India; *v.rajinikanth@ieee.org*

[2] Department of Information Technology, Techno India College of Technology, Kolkata-700156, West Bengal, India; *neelanjan.dey@gmail.com*

[3] Key Laboratory of Digital Signal and Image Processing of Guangdong Province, Department of Electronic Engineering, College of Engineering, Shantou University, Shantou, Guangdong, China; *jalexnoel@stu.edu.cn*

[4] Faculty of Computers and Artificial Intelligence, Cairo University, Egypt; *aboitcairo@cu.edu.eg*

[5] Department of Computer Science, University of South Dakota, 414 E Clark St, Vermillion, SD, 57069, USA; *santosh.kc@ieee.org*

**Abstract:** Pneumonia is one of the foremost lung diseases and untreated pneumonia will lead to serious threats for all age groups. The proposed work aims to extract and evaluate the Coronavirus disease (COVID-19) caused pneumonia infection in lung using CT scans. We propose an image-assisted system to extract COVID-19 infected sections from lung CT scans (coronal view). It includes following steps: (i) Threshold filter to extract the lung region by eliminating possible artifacts; (ii) Image enhancement using Harmony-Search-Optimization and Otsu thresholding; (iii) Image segmentation to extract infected region(s); and (iv) Region-of-interest (ROI) extraction (features) from binary image to compute level of severity. The features that are extracted from ROI are then employed to identify the pixel ratio between the lung and infection sections to identify infection level of severity. The primary objective of the tool is to assist the pulmonologist not only to detect but also to help plan treatment process. As a consequence, for mass screening processing, it will help prevent diagnostic burden.

**Keywords:** Lung abnormality; CT scans; COVID-19 infection; Pneumonia; Image processing system; Severity detection.

## 1. Introduction

The increased medical facility in the modern era is committed to provide a better living atmosphere to the human community by preventing life-threatening diseases. Due to various uncontrolled and unavoidable reasons, communicable and chronic diseases in humans are gradually rising globally. Most of these diseases can be controlled and completely cured when it is identified in its premature stage. When disease symptom is experienced, the patient approaches the hospital for the diagnosis to identify the cause, nature and the severity of the disease.

The disease in the internal body organs are very acute compared to other illnesses and hence, more diagnostic methods are recommended and implemented to identify the disease and its severity to plan for an appropriate treatment procedure to control the disease. In human physiology, the lung is one of the prime organs responsible to circulate the air and the infection/disease in lung will severely affect the respiratory system and the untreated lung disease will lead to death. The major causes of lung defects are due to diseases, such as tuberculosis, lung-cancer, and pneumonia [1,2]. Further, the infection in lungs due to microorganisms, such as bacteria, virus and fungi may cause mild to harsh illness; which may require instantaneous medical support to cure the disease.

When a patient is admitted to the hospital due to respiratory infection, the pulmonologist will recommend a series of diagnostic procedures ranging from the non-laboratory to laboratory tests to identify the cause, location, and severity of the infection. The laboratory tests include common procedures, like complete-blood-count, blood-gas-analysis and pleural-fluid-analysis [3,4]. The non-laboratory tests are mainly the image assisted techniques used to record or inspect the lung regions using chest radiography (X-ray), Computed-Tomography (CT) scan and bronchoscope. Due to its non-invasive nature, the chest X-ray and CT are commonly used to record the digital images of the lungs, which can be further examined by an experienced doctor or a computer-supported system to identify the severity of the infection. Compared to the chest X-ray, the CT Scan Images (CTSI) is widely preferred due to its merit and the three-dimensional view of the lung; which further can be used to provide 2D images of axial, coronal and Sagittal-View for the better diagnosis. The recent studies also confirmed the merit of the CT in detecting the infection in the respiratory tract and pneumonia [5].

The objective of the proposed work is to develop an image processing procedure to extract the infected section of COVID-19 from lung CTSI. The recent studies on COVID-19 infection confirmed that, the assessment of disease severity rate is essential to plan for the appropriate treatment to cure the patients. In real-time scenarios, infected regions from the bio-imaging methods, such as CT and chest radiographs are assessed by an experienced doctor and based on the recommendation; possible treatment procedures are suggested and implemented to reduce the infection level. The personal detection with an experienced doctor will be a tedious task when the number of patients is large and hence, it is necessary to develop an automated/semi-automated image processing procedure to assist the doctor to examine the lung images of the patients. This procedure will considerably reduce the diagnostic burden of the doctor by segregating the lung CTSI based on the severity level. Further, this procedure also will assist the doctor to track the changes in infection rate (reduction/progress in infection) when the treatment is implemented.

The current literature on COVID-19 confirms that the disease infection rate in a patient is normally performed with lung CT/X-ray images and correct detection of the infection level is very essential to plan for an appropriate treatment process. The infection visibility in chest X-ray is very poor and it needs the expertise to identify and understand the information; hence in most of the recent works; CTSI is preferred compared to the chest X-ray. Further, the work of Li et al. [6] also suggested that, visual quantitative analysis using the CTSI provided good reliability and better diagnostic capability. Hence, the proposed work considered the coronal-view of the two-dimensional CTSI for the experimental investigation. The proposed study is investigated in the MATLAB® environment and the essential CTSI of COVID-19 is collected from the Radiopedia database [7].

This work implements a series of procedures, such as threshold filtering, multi-thresholding, segmentation and area feature extraction to detect the COVID-19 pneumonia infection from the considered CTSI. After extracting the infected region with better accuracy, the infection stage is then computed using the ratio of pixels in the infection/lung section. This Severity-Level-Measure (SLM) is then considered to short the existing patient's CTSI and the image that has a higher value of SLM will be sent to the pulmonologist for the approval. The pulmonologist will also perform a visual check on the CTSI, and as per the findings, the appropriate treatment will be initiated.

The remaining part of this research is presented as follows; section 2 discusses the existing works on the COVID-19 database, section 3 outlines the methodology implemented, sections 4 and 5 present the experimental outcome and the conclusion, respectively.

## 2. Related Works on COVID-19 Infection

COVID-19 is an infectious disease caused by Severe Acute Respiratory Syndrome-Corona Virus-2 (SARS-CoV-2) and this sickness is first revealed in China (Wuhan) in December 2019 [7, 8]. Due to the outbreak, COVID-19 has emerged as a global challenge and a significant number of research works are initiated to discover the solution to control the infection rate and spreading of the COVID-19 [9].

Due to its significance, a significant amount of research works are proposed to assess pneumonia caused by COVID-19 and the severity of this infection is to be accounted to take the decision regarding; (i) Treatment procedure to be followed and (ii) The choice of the drug and its dosage level. The non-invasive image assisted detection procedures are followed in the hospitals to detect the severity of the COVID-19 pneumonia and the imaging modalities, such as CT and Chest X-ray are widely considered. The information seen in CT is more clear compared to the Chest X-ray and hence, this research work considered only the CTSI for the examination, Moreover, the resent findings by [10] confirmed that the CT offers better judgment accuracy compared to the laboratory level detection process called Reverse transcription-polymerase Chain Reaction (RT-PCR). The recent procedures to detect COVID-19 pneumonia using the CTSI are summarized in Table 1.

**Table 1.** Summary: recently reported COVID-19 Pneumonia detection procedure.

| Reference | Procedure | Findings |
|---|---|---|
| Li et al. [6] | This work proposed a methodology to identify the infection rate in 78 patients (38 men and 40 women) using the axial and coronal-view of lung CTSI. | The proposed work helped to attain better performance values;, such as AUC (91.8%), sensitivity (82.6%) and specificity (100%). |
| Chung et al. [9] | This work presented a visual inspection based evaluation of COVID-19 disease using two-dimensional axial-view CTSI collected from 21 patients from China with age ranging from 29-77. This work confirmed that, visual inspection helps to identify the infection with greater accuracy. | This work identified the severity by computing the overall lung total severity score (LTSS). |
| Bernheim et al. [11] | This work proposed an analysis using axial-view CTSI to identify the COVID-19 infection in 121 symptomatic patients from China; categorized as early (0-2 days), intermediate (3-5 days) and late (6-12 days) are collected and a visual check with experienced radiologist is carried to confirm the disease. | The experimental investigation confirmed that the infection pattern of COVID-19 is approximately similar with the infection caused by SARS and MERS. |
| Yan et al. [12] | This work proposed an evaluation using the lung CTSI of axial-view and this work considered the images of 102 volunteers (53 men and 49 women with age group of 15-79 years). This work also implemented visual inspection based detection. | Implemented detection helped to attain Area-Under-Curve (AUC) of 89.2%, Sensitivity of 83.3% and Specificity of 94.3%. |
| Wang et al. [13] | This work examined the lung CTSI of 90 patients (33 men and 57 women with a mean age of 45 years). This work collected 366 number of CT slices and examined by two group of experienced radiologists. | The examination outcome confirmed that, the infection will be severe during 6-11 days from the |

| | | infection. This work also helps to identify the progression of disease based on number of days. |
|---|---|---|
| Shi et al. [14] | This work implemented an image feature and laboratory assisted diagnostic procedure to detect the COVID-19 infection using the CTSI of 81 patients (42 men and 39 women with a mean age of 49.5 years). The findings of this work confirmed that, the infection rate is severe during 1-3 weeks. | This work confirmed that, the combined evaluation of imaging features along with clinical and laboratory findings will assist in premature detection of pneumonia caused by COVID-19. |
| Fang, et al. [15] | This work examined the information of 81 patients for the examination. Among the 81 patients, 30 patients are evaluated with RT-PCR and 51 patients are evaluated with both CTSI and RT-PCR. This work combined the laboratory analysis (RT-PCR) along with the imaging procedure based on the CTSI. | The sensitivity of CTSI based detection of COVID-19 infection was 98% contrast to RT-PCR sensitivity of 71% (p<.001). |
| Bai et al. [16] | This work examined 219 patient's information using the radiologist team of China and U.S and their findings are discussed. This work considered the RT-PCR and CTSI for the examination and in this work. This work also presented a classification task using 219 COVID-19 along with 205 non-COVID-19 pneumonia case. | The experimental findings presented a better sensitivity (93%) and specificity (100%) for both the China and U.S radiologist. |
| Chua et al. [17] | This work presented a clinical evaluation of 36 patients scanned with the first two days of symptoms. These patients are evaluated with CTSI as well as real-time RT-PCR and the results confirmed that, the early detection accuracy of RT-PCR (>90%) is better compared to CTSI (56%). | COVID-19 infection is highly transmissible with a relatively low death rate (1·0–3·5%) for individuals with lower age (< 70 years). |
| Liu et al. [18] | This work presented a procedure to detect the infection using RT-PCR and CTSI information of 73 patients. This work also suggested the implementation and follow-up procedures of the combined assessment of RT-PCR and CTSI. | The proposed work classified the patient's condition, like mile (6 patients), common (43 patients), severe (21), and critical (3). |
| Zhou et al. [19] | This work evaluated the information of 62 patients (34 men and 28 women of ages 20-91). This work implemented the examination using the axial and coronal view of the non-contrast CTSI. | The proposed analysis confirmed that the CT assisted evaluation shows better detection accuracy in progressive stage confirmed to the early stage. |
| Yoon et al. [20] | This work proposed an image assisted examination procedure for COVID-19 using the CTSI and Chest X-ray images. This work evaluated 9 patient's data (4 men and 5 women with a median age of 54 years) and the work | This research confirmed that the radiology image of COVID-19 pneumonia is milder compared to the |

| | confirms that, image based assessment can be used to detect the COVID-19 infection with better accuracy. | pneumonia caused by SARS and MERS. |
|---|---|---|

Due to its significance, recently researchers proposed a considerable number of COVID-19 prediction and detection procedures [21-23]. The recent work of Rodriguez-Morales et al. [24] presents a detailed review of the detection and prediction of COVID-19 pneumonia infection. The existing identification procedures involves in laboratory level detection using RT-PCR and physical assessment of CTSI using an experienced radiologist. The computerized detection procedure is essential to reduce the assessment burden of the radiologist and support the fast and accurate decision-making process.

The proposed research work aims to develop an image-assisted procedure to detect and extract the COVID-19 infected section from the CTSI. This work initially considered the coronal-view of the CTSI for the assessment and later, a similar procedure is executed to evaluate the axial-view of CTSI. The proposed work also proposed a methodology to detect the average infection rate by considering the pixel information of the infected section and the lung section. The ratio between the pixels of the infected and lung segments is considered to examine the severity of the COVID-19 pneumonia.

## 3. Materials and Methods

This part of the study discusses the proposed method implemented to detect the pneumonia infection from CTSI. The various stages involved in the proposed technique are presented in Figure 1. Initially, the two-dimensional CTSI is collected from the benchmark database. The CTSI is associated with the lung section along with the artifacts, such as the bone and other body sections. The threshold filler discussed in Bhandary et al. [4] is then implemented to separate the lung from the artifact. Later, a multi-threshold based on Otsu function and Harmony-Search-Optimization (HSO) is implemented to enhance the visibility of the infected lung segment. The enhanced section is then extracted using watershed segmentation. Finally, the pixel values of the binary images of infected and the lung region are considered to compute the severity rate.

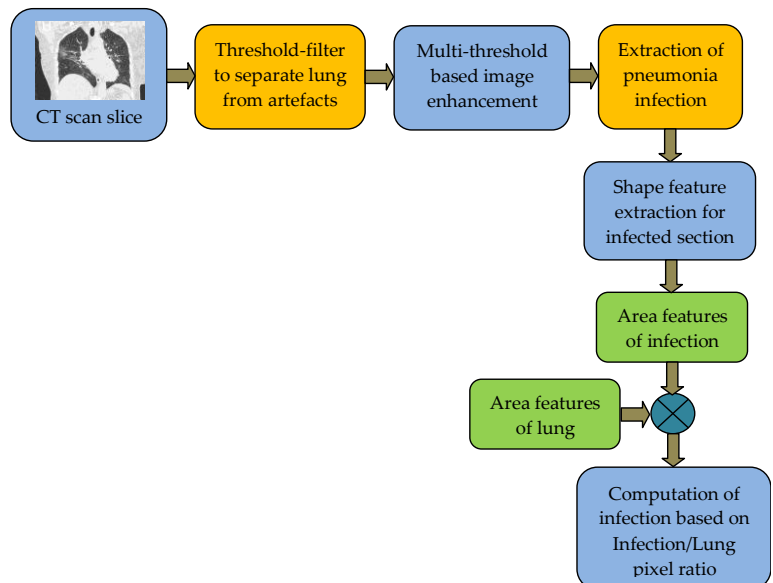

**Figure 1.** Stages involved in the proposed automated infection detection system

*3.1 COVID-19 database*

The COVID-19 is a new class of infection, which causes mild to severe pneumonia according to the infection rate and the age of the individuals [5, 18-20]. The existing clinical level detection of the infection involves laboratory analysis with RT-PCR and imaging-based detection with CTSI and chest X-ray. Due to various restrictions, the collection of the radiology images from patients is a challenging task; further, the COVID-19 is a new category of disease and hence the image database existing is very limited.

The proposed work considered the images of the Radiopedia database for the evaluation [7]. Radiopedia shared the clinical-grade chest X-ray and CTSI for the research purpose. This work used the clinical-grade CTSI of coronal-view images of case studies, such as 1 [25], 4 [26], 5 [27], 7 [28], 15 [29], 38 [30], 39 [31], and 45 [32] for the experimental investigation. Along with the above said images, CTSI of case study 30 [33] is also considered to test the performance of the proposed procedure on the axial-view images. From each case study, ten numbers of 2D CTSI are extracted and examined. Further, this work also considered the COVID-19 images available in [34] for the examination. This dataset consist the axial and coronal-view of the CTSI recorded with and without the contrast agent.

This work considered 90 2D slices of coronal-view (9 patients x 10slices = 90 slices) and 20 2D slices of axial-view (2 patients x 10slices = 20slices) CTSI for the experimental investigation. The sample test images of the coronal-view CTSI is depicted in figure 2 and the dimension of the images are considered as given in the Radiopedia database.

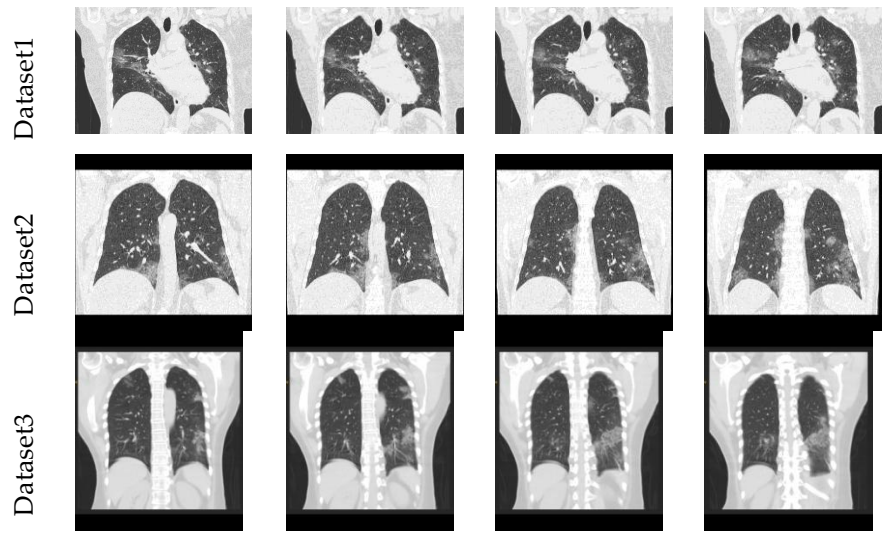

**Figure 2.** Few test images of COVID-19 dataset offered by Radiopedia

*3.2 Threshold filter*

The accuracy in the medical image evaluation depends mainly on the images considered for the examination. In this work, the artifacts (bone and other body segments) existing in the considered test image is initially removed using the threshold filter discussed in [4]. The concept of this filter is simple and it separates the test image into two sections based on the chosen threshold value. In this work, the threshold values are selected manually by trial and error analysis and the chosen threshold then works well on all the images considered in the proposed study.

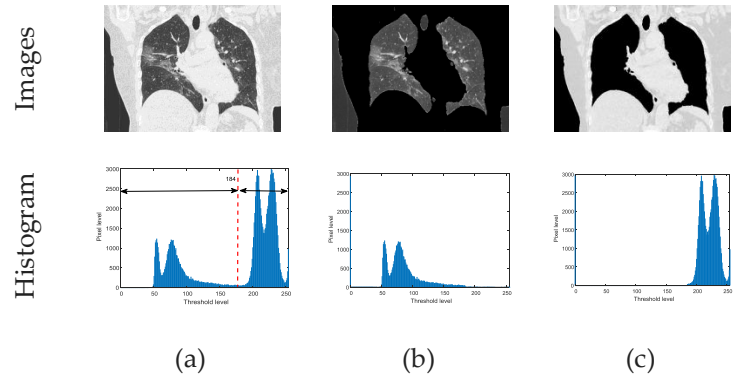

**Figure 3.** Outcomes of the threshold filter: (a) Test image, (b) Lung segment, and (c) Artifact.

The considered images can be exactly separated based on its pixels, when the threshold is chosen as 184 (i.e. < 184 = lung and >184 is artifact). The results attained with the threshold filter are depicted in figure 3. Fig 3(a) depicts the original test image and the gray histogram; and Fig 3(b) and (c) depict the image and histograms of the lung and the artifacts, respectively. The similar threshold values are considered for other test images, in this study.

*3.3 Image enhancement*

In the literature, a number of medical image enhancement procedures are available to pre-process the test images. This work employed the three-level threshold-based enhancement of medical imaging procedure discussed in the recent works [35,36]. This pre-processing procedure will enhance the test image by grouping alike pixels of the background, normal lung fragment and the COVID-19 infection. In this work, Harmony-Search-Optimization (HSO) algorithm is implemented to find the optimal threshold due to its simplicity and lesser tuning parameters. Initially, the threshold selection is done using the entropy techniques [35,36] and Otsu between-class function [37] and the enhancement achieved with Otsu is superior compared to other technique. Hence, the entire work is executed using the HSO + Otsu. The essential information on the thresholding can be found in the earlier works of Oliva et al. [38] and Cuevas et al. [39]. This work considered the HSO+Otsu based threshold implemented in [38-40]. Other information on Otsu thresholding can be found in the earlier research works available in [41].

*3.4 Segmentation of Pneumonia Infection*

In the medical image examination problem, the extraction of the abnormal section from the test image is essential for further assessment. The extraction of an abnormal section is normally done using a chosen image segmentation procedure and, in this work, an automated segmentation procedure called the watershed technique is implemented. The watershed approach implements a series of techniques, such as edge detection, morphological operation, image-fill, enhancement, and extraction. The recent work of Bhandary et al. [4] confirms that the watershed technique helped to extract the abnormal section from the CTSI with better accuracy. The extracted abnormal section is in the form of the binary in which the background is assigned with a value '0' and the extracted region is assigned with '1' and this section is considered for further assessment, if necessary.

## 3.5 Severity Level Computation

The hospital level examination of the COVID-19 is performed with a laboratory test: RT-PCR to confirm the infection. After admitting the patient, the imaging procedures, such as CT/X-ray data are considered to identify infected section and its level of severity. Based on the severity level, the appropriate treatment procedure is to be planned which involves in the need for the assisting devices, monitoring devices, and the choice of the drug and its dosage level. Further, the imaging procedure also assists the pulmonologist to confirm the recovery of the patient with respect to the treatment currently implemented. For each patient, the imaging procedure is to be followed for a considerable number of times to record and confirm the recovery. A personal analysis of the CTSI with an experienced radiologist is essential for an accurate diagnosis. The increase in patient's rate will improve the diagnostic burden of the radiologist and hence a computer-supported assisting methodology is essential to get the preliminary diagnosis regarding the pneumonia infection. In this work, an image assisted disease severity level is proposed to identify the infection severity based on the mathematical expression depicted in Eqn. 1;

$$Percentage\ infection\ rate = \frac{Pixel\ density\ in\ extracted\ section}{Pixel\ density\ of\ the\ lung\ segment} \times 100 \qquad (1)$$

In this work, the binary image of the lung can be extracted using ITK-Snap [42,43], an open-access software available for the assessment of medical images. If this tool is implemented, a lab technician can perform the proposed image examination task and the experimental outcome along with the considered CTSI is then submitted for the suggestion of an experienced radiologist.

## 4. Result and Discussions

This part presents the experimental result obtained from the proposed system and its dissuasion. The proposed work is implemented using a workstation with the Intel i5 processor with a clock frequency 2.GHz with 8GB RAM and 2DB VRAM. The experimental outcome confirms that the proposed system takes a mean time of 38 sec to process a patient's data (10 slices) and provides the mean value of infection rate for further evaluation.

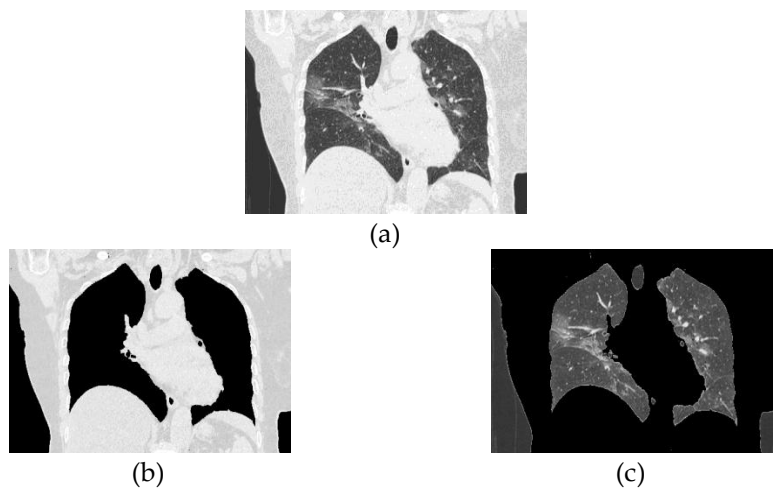

(a)

(b)　　　　　　　　　　　　(c)

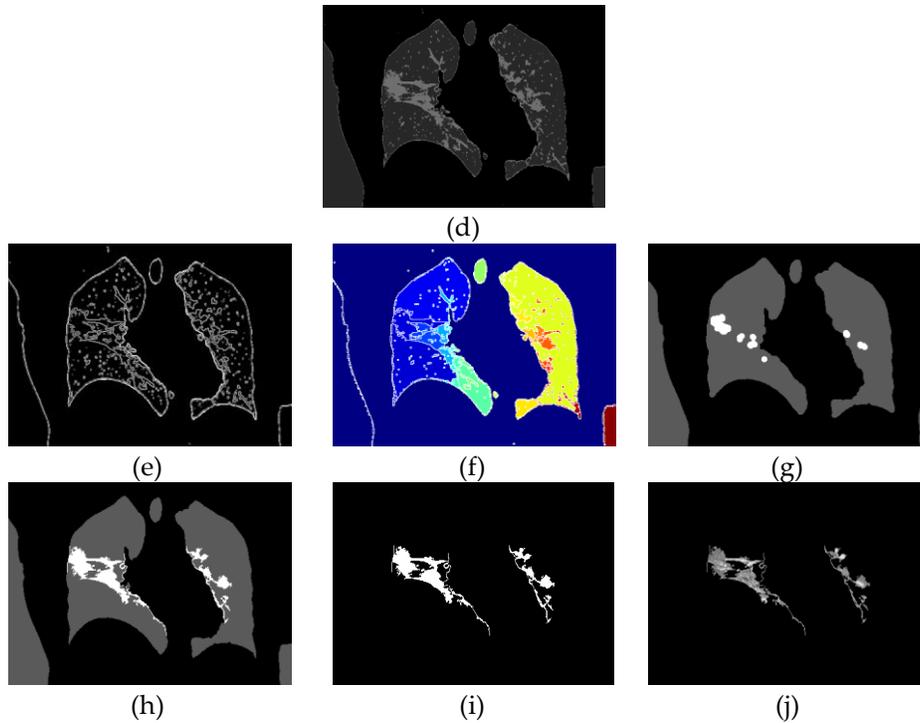

**Figure 4.** Results obtained using the proposed automated pneumonia detection system:
(a) Test image considered; (b) Extracted artifact part; (c) Extracted lung section;
(d) Thresholding result; (e) Detected edges; (f) Watershed fil;, (g) Morphological enhancement; (h) Enhancement of infection; (i) Binary format of infected section;
(j) Extracted infection.

Figure 4 depicts the experimental results attained for the chosen trial CTSI. Fig 4(a) presents the coronal-view test image considered for the demonstration. Fig 4(b) and (c) presents the threshold filter outcomes, like artifacts and lung section, respectively. Fig 4(d) shows the thresholded picture with HSO + Otsu and Fig 4(e) – (g) depicts the intermediate results of the watershed algorithm. Fig 4(h) represents the enhanced infection and Fig 4(i) depicts the extracted binary form of the infection. Finally, the infection available in the trial image is presented in Fig 4(j), which is attained by implementing the pixel level multiplication of Fig 4(a) and Fig 4(i). Similar results are attained for other images considered in this research work.

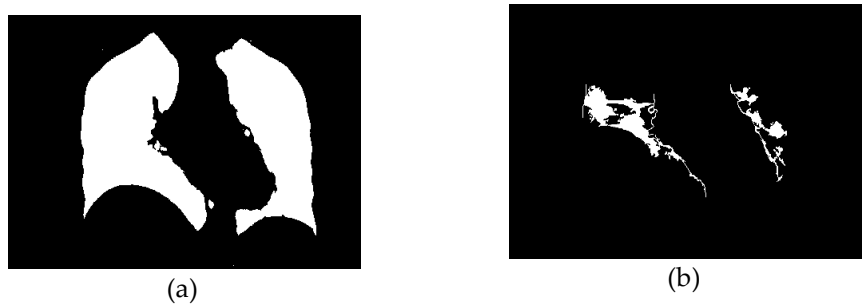

**Figure 5.** (a) Lung with a pixel density of 20322 and (b) Infection with a pixel density of 2948

After extracting the infected section from the trial picture, its infection rate is computed as discussed in Eqn. 1. To achieve this, the lung section from this image is extracted with ITK-Snap and then the binary pixels of the lung section as well as the extracted section are computed. The outcome of this procedure is depicted in figure 5 and the results attained for lung and COVID-19 infection are presented in Fig 5(a) and (b) respectively. The infection rate existing in the considered image is computed as follows;

$$Infection\,rate\,of\,this\,image = \frac{2948}{20322} \times 100 = 14.51\% \qquad (2)$$

Similar procedure is implemented for other nine slices of this case study and for this patient; the mean infection rate of 14.17% is achieved.

This examination procedure is then considered to evaluate the COVID-19 infection in other CTSI of the case studies and the results are noted. The sample experimental outcome for the coronal and axial-view images is depicted in figure 6. In which, Fig 6(a)-(c) depicts the chosen test images and the threshold filter results. Fig 6(d) shows the thresholded image. Fig 6(e) and (f) denote the outcome of the watershed algorithm. Fig 6(g) and (h) present the segmented binary image and the lung infection respectively.

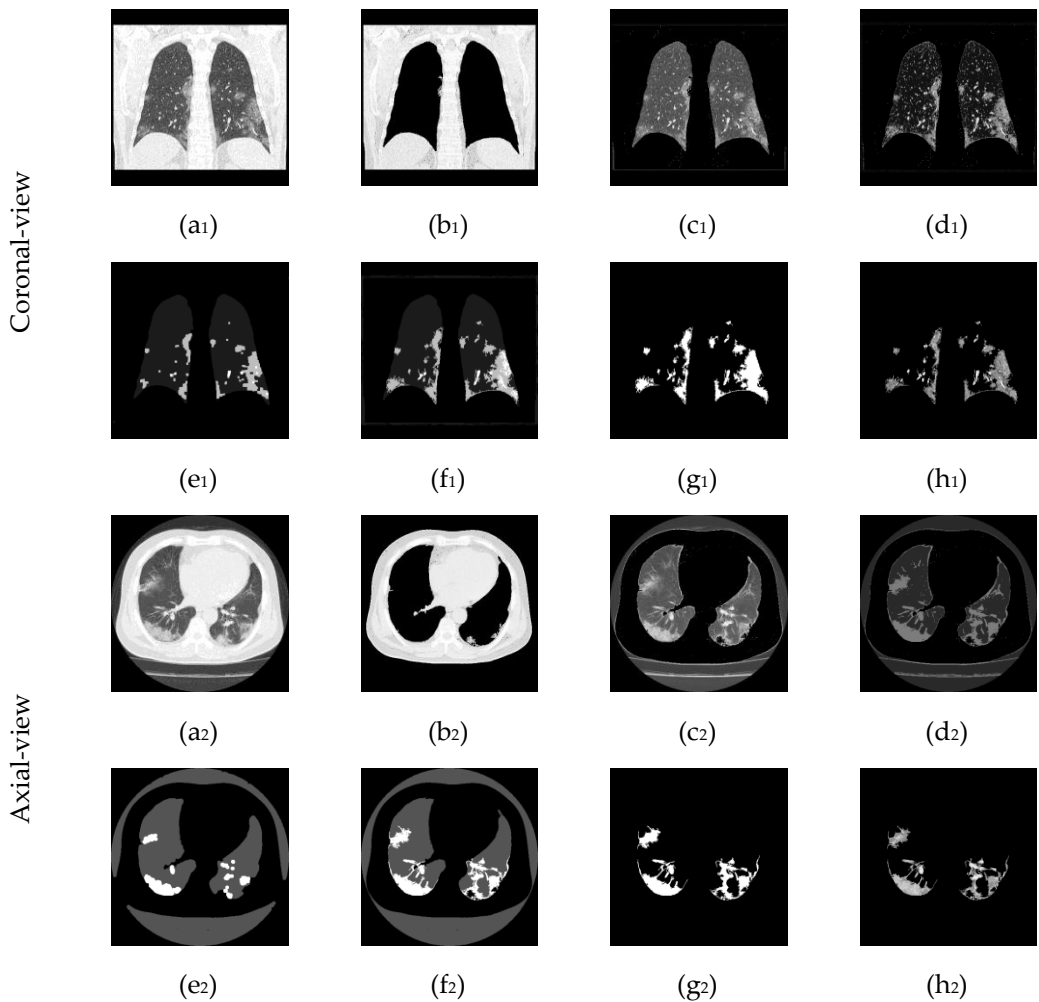

**Figure 6.** Results attained for coronal and axial view test images of COVID-19 infected lung CT Images
(Subscripts; 1 and 2 denote the coronal and axial-view slices, respectively)

The proposed image-assisted detection procedure works well on the CTSI of the coronal and axial-view and helps achieve a better segmentation of the infected lung section. The proposed technique offered the following average infection rate for the considered test images: Case1 = 15.96%, Case4 = 18.24%, Case5 = 19.05%, Case7 = 14.17%, Case15 = 32.02%, Case30 = 16.21%, Case38 = 13.85%, Case39 =

9.95%, and Case45 = 17.58%. The patient with a larger infection rate will get more priority for the treatment compared to the patient with a lesser rate. Further, this analysis also can be used to detect the progress of the COVID-19 infection in a patient; using the CTSI recorded with various time intervals. This measure can also be considered to detect the recovery rate of the patient.

Along with the above said patient cases, this work also considered the COVID-19 pneumonia infection dataset of 75-year-old (male) patient for the assessment [27,34] and the experimental investigation obtained with axial-view (10 slices) and coronal-view (10 slices) lung images offered an average infection rate of 38.52% and 39.06% respectively. This result confirmed that, proposed work helps to attain a better detection of the lung infection irrespective of the CTSI slice orientation (i.e. coronal/axial). Further, the infection rates computed for both the image cases are approximately similar, which confirms the clinically significance of proposed technique.

This work proposed preliminary research to examine the COVID-19 infection using the CTSI of the coronal and axial-view dataset. The COVID-19 is a novel disease and hence the collection of clinical-grade images is a difficult task. This work executed a series of procedure to extract the infected section from the 2D lung CT scan slices and also proposed a mathematical expression to compute the infection rate using a pixel-level assessment. In future, this work can be extended in the following directions: (i) Developing a procedure to detect and categorize the COVID-19 infection mild, moderate, and severe class; (ii) Image assisted automated follow-up study to detect the progress of the disease; and (iii) Implementing an automated detection system to classify the existing CTSI slices into normal/COVID-19 pneumonia class.

## 5. Conclusion

The main aim of the proposed research work is to develop an image assisted examination procedure to extract and assess the COVID-19 pneumonia infection from the lung CT scan images (CTSI). This work proposed a series of procedures to extract the infected section from the considered CTSI with images orientation, such as coronal and axial. This work also suggests a procedure to compute the infection rate in the lung based on the pixel level ratio of the infected and lung region. This work considered the clinical-grade 2D slices of the CTSI for the examination and the result attained in this study confirms that proposed procedure is efficient in extracting the infected section with better accuracy. Further, this infection rate can be used to classify the considered test images into various classes based on the amplitude of the infection rate.

Considering these preliminary results, our immediate plan is to examine COVID-19 benchmark datasets and clinical study (collected from hospitals). The proposed system, however, is currently, not suitable to examine chest X-Ray images of the patients that are COVID-19 infected.

**Funding:** This research received no external funding

**Acknowledgments:** The authors of this research would like to thank Radiopedia.org for sharing the clinical grade COVID-19 images.

**Conflicts of Interest:** The authors declare no conflict of interest.

## References

1. Syrjala, H. et al. Chest magnetic resonance imaging for pneumonia diagnosis in outpatients with lower respiratory tract infection. European Respiratory Journal 2017, 49, 1601303. DOI: 10.1183/13993003.01303-2016
2. Zech, JR. et al. Variable generalization performance of a deep learning model to detect pneumonia in chest radiographs: a cross-sectional study. PLoS Med. 2018, 15 (11), Article p.e1002683.

3. Rajpurkar, P. et al., CheXNet: radiologist-level pneumonia detection on chest x-rays with deep learning. (2017). arXiv:1711.05225 [cs.CV].
4. Bhandary, A. et al. Deep-learning framework to detect lung abnormality–A study with chest X-Ray and lung CT scan images. Pattern Recogn Lett 2020, 129, 271-278. https://doi.org/10.1016/j.patrec.2019.11.013.
5. Nascimento, IBD, et al. Novel Coronavirus Infection (COVID-19) in Humans: A Scoping Review and Meta-Analysis. J. Clin. Med. 2020, 9(4), 941; https://doi.org/10.3390/jcm9040941
6. Li, K., Fang, Y., Li, W. et al. CT image visual quantitative evaluation and clinical classification of coronavirus disease (COVID-19). Eur Radiol (2020). https://doi.org/10.1007/s00330-020-06817-6.
7. https://radiopaedia.org/articles/COVID-19-3?lang=us (Last accessed date 5th April 2020)
8. Song, F, Shi, N, Shan, F, Zhang, Z, Shen, J, Lu, H et al. Emerging Coronavirus 2019-nCoV Pneumonia. Radiology 2020, 295, 210–217.
9. Chung, M, Bernheim, A, Mei, X, Zhang, N, Huang, M, Zeng, X et al. CT Imaging Features of 2019 Novel Coronavirus (2019-nCoV). Radiology 2020, 295, 202–207. https://doi.org/10.1148/radiol.2020200230.
10. https://healthcare-in-europe.com/en/news/ct-outperforms-lab-diagnosis-for-coronavirus-infection.html (Last accessed date 5th April 2020)
11. Bernheim, A. et al. Chest CT Findings in Coronavirus Disease-19 (COVID-19): Relationship to Duration of Infection. Radiology 2020. https://doi.org/10.1148/radiol.2020200463
12. Yan, R. et al. Chest CT Severity Score: An Imaging Tool for Assessing Severe COVID-19. Radiology: Cardiothoracic Imaging 2020, 2(2). https://doi.org/10.1148/ryct.2020200047.
13. Wang, Y. et al. Temporal Changes of CT Findings in 90 Patients with COVID-19 Pneumonia: A Longitudinal Study. Thoracic Imaging 2020. https://doi.org/10.1148/radiol.2020200843.
14. Shi, H. et al. Radiological findings from 81 patients with COVID-19 pneumonia in Wuhan, China: a descriptive study. Lancet Infect Dis. 2020, 20(4), 425-434. https://doi.org/10.1016/S1473-3099(20)30086-4.
15. Fang, Y. et al. Sensitivity of chest CT for COVID-19: comparison to RT-PCR. Radiology 2020. DOI:10.1148/radiol.2020200432.
16. Bai, H.X. et al. Performance of radiologists in differentiating COVID-19 from viral pneumonia on chest CT. Radiology 2020. DOI: 10.1148/radiol.2020200823.
17. Chua, F. et al. The role of CT in case ascertainment and management of COVID-19 pneumonia in the UK: insights from high-incidence regions. Lancet Resp Med 2020. https://doi.org/10.1016/S2213-2600(20)30132-6.
18. Liu, K-C. et al. CT manifestations of coronavirus disease-2019: A retrospective analysis of 73 cases by disease severity. Eur J Radiol 2020, 126, 108941. https://doi.org/10.1016/j.ejrad.2020.108941.
19. Zhou, Z., Guo, D., Li, C. et al. Coronavirus disease 2019: initial chest CT findings. Eur Radiol (2020). https://doi.org/10.1007/s00330-020-06816-7.
20. Yoon, SH. et al. Chest Radiographic and CT Findings of the 2019 Novel Coronavirus Disease (COVID-19): Analysis of Nine Patients Treated in Korea. Korean J Radiol. 2020, 21(4):494-500. Doi: 10.3348/kjr.2020.0132.
21. Fong, SJ., Li, G., Dey, N., Crespo, R.G., Herrera-Viedma, E. Finding an Accurate Early Forecasting Model from Small Dataset: A Case of 2019-nCoV Novel Coronavirus Outbreak. International Journal of Interactive Multimedia and Artificial Intelligence 2020, 6(1), 132-140. Doi: 10.9781/ijimai.2020.02.002.
22. Fong, SJ., Li, G., Dey, N., Crespo, R.G., Herrera-Viedma, E. Composite Monte Carlo Decision Making under High Uncertainty of Novel Coronavirus Epidemic Using Hybridized Deep Learning and Fuzzy Rule Induction. 2020, 9. arXiv:2003.09868 [cs.AI].
23. Santosh, KC. AI-Driven Tools for Coronavirus Outbreak: Need of Active Learning and Cross-Population Train/Test Models on Multitudinal/Multimodal Data. J Med Syst 2020, 44, 93. https://doi.org/10.1007/s10916-020-01562-1.

24. Rodriguez-Morales, AJ. et al. Clinical, laboratory and imaging features of COVID-19: A systematic review and meta-analysis. Travel Medicine and Infectious Disease 2020, 101623. https://doi.org/10.1016/j.tmaid.2020.101623.
25. Case courtesy of Dr Chong Keng Sang, Sam, Radiopaedia.org, rID: 73893 (Last accessed date 5th April 2020)
26. Case courtesy of Dr Domenico Nicoletti, Radiopaedia.org, rID: 74724 (Last accessed date 5th April 2020)
27. Case courtesy of Dr Fabio Macori, Radiopaedia.org, rID: 74867 (Last accessed date 5th April 2020)
28. Case courtesy of Dr Fateme Hosseinabadi , Radiopaedia.org, rID: 74868 (Last accessed date 5th April 2020)
29. Case courtesy of Dr Derek Smith, Radiopaedia.org, rID: 75249 (Last accessed date 5th April 2020)
30. Case courtesy of Dr Bahman Rasuli, Radiopaedia.org, rID: 74880 (Last accessed date 5th April 2020)
31. Case courtesy of Dr Mohammad Taghi Niknejad, Radiopaedia.org, rID: 75605 (Last accessed date 5th April 2020)
32. Case courtesy of Dr Mohammad Taghi Niknejad, Radiopaedia.org, rID: 75662 (Last accessed date 5th April 2020)
33. Case courtesy of Dr Bahman Rasuli, Radiopaedia.org, rID: 75330 (Last accessed date 5th April 2020)
34. https://radiopaedia.org/cases/COVID-19-pneumonia-12 (Last accessed date 5th April 2020)
35. Fernandes, SL, Rajinikanth, V, Kadry, S. A hybrid framework to evaluate breast abnormality using infrared thermal images. IEEE Consum. Electron. Mag 2019,  8(5), 31–36. https://doi.org/10.1109/MCE.2019.2923926.
36. Dey, N. et al. Social-group-optimization based tumor evaluation tool for clinical brain MRI of Flair/diffusion-weighted modality. Biocybern. Biomed. Eng. 2019, 39(3),843–856. https://doi.org/10.1016/j.bbe.2019.07.005
37. Satapathy SC, Raja NSM, Rajinikanth V, Ashour AS, Dey N. Multi-level image thresholding using Otsu and chaotic bat algorithm. Neural Comput Appl 2018, 29(12),1285–1307. https://doi.org/10.1007/s00521-016-2645-5.
38. Oliva, D. et al. Multilevel thresholding segmentation based on harmony search optimization. Journal of Applied Mathematics 2013, 2013, 575414.  https://doi.org/10.1155/2013/575414.
39. Cuevas, E. et al. Otsu and Kapur Segmentation Based on Harmony Search Optimization.  Intelligent Systems Reference Library 2015, 100, 169-202.
40. Geem ZW, Kim JH, Loganathan GV. A new heuristic optimization algorithm: harmony search. Simulations 2001, 76, 60–68. https://doi.org/10.1007/978-3-319-26462-2_8.
41. Otsu, N. A threshold selection method from gray-level histograms. IEEE Trans. Syst. Man Cybern 1979, SMC-9, 62–66.
42. http://www.itksnap.org/pmwiki/pmwiki.php (Last accessed date 5th April 2020)
43. aul A. Yushkevich, Joseph Piven, Heather Cody Hazlett, Rachel Gimpel Smith, Sean Ho, James C. Gee, and Guido Gerig. User-guided 3D active contour segmentation of anatomical structures: Significantly improved efficiency and reliability. *Neuroimage* 2006, 31(3),1116-1128.